\documentclass[a4paper,12pt]{article}
\usepackage{graphicx}
\usepackage{epsfig}
\usepackage{jcappub}
\newcommand{\be}{\begin{equation}}
\newcommand{\ee}{\end{equation}}

\title{GCG Parametrization for Growth Function and Current Constraints}
\date{\today}

\author{Gaveshna Gupta, Somasri Sen\footnote{Presently at:\\ Department of Physics\\Jamia Millia Islamia, New Delhi\\
110025, India} and Anjan A Sen}
\affiliation{Center For Theoretical Physics, Jamia Millia Islamia, New Delhi-110025, India}

\emailAdd{gaveshna@ctp-jamia.res.in, ssen@jmi.ac.in, aasen@jmi.ac.in}

\abstract{  We study the linear growth function $f$ for large scale structures in a cosmological scenario where Generalised Chaplygin Gas (GCG) serves as dark energy candidate. 
We parametrize the growth index parameter as a function of redshift and do a comparative study  between the theoretical growth rate and the proposed parametrization. Moreover, we  demonstrate that growth rates for a wide range of dark energy models  can be modeled accurately by our proposed parametrization.  Finally,  we compile a data set consisting of 28 data points within redshift range (0.15,3.8) to constrain the growth rate. It includes direct growth data from various projects/surveys including the latest data from the  Wiggle-Z measurements. It also includes data constraining growth indirectly through the rms mass fluctuation $\sigma_8(z)$ inferred from Ly-$\alpha$ measurements at various redshifts. By fitting our proposed parametrization for $f$ to these data, we show that growth history of large scale structures of the universe although allows a transient acceleration, one cannot distinguish it at present with an eternally accelerating universe.
}


\begin{document}

\maketitle

\section{Introduction}

The accelerating phase of the present cosmic expansion is attributed to 
an ideal fluid dubbed {\em Dark energy} \cite{Bean} \cite{copeland} \cite{Padmanabhan} \cite{Peebles} \cite{Sahni}. The evidences for the existence of such dark entity come mainly from the observations that measure the expansion rate of the universe $ H(a)\equiv \dot{a}/{a} $ at different epoch, where $ a $ is the scale factor of the universe. Some of the important tests which provide most precise probes for dark energy by measuring the expansion rate of the universe are - the  luminosity distance measurement from standard candles like Type Ia Supernovae \cite{Riess1998} \cite{Perlmutter1999} \cite{Tonry2003} \cite{Knop2003} \cite{Riess2004}, the measurement of angular diameter distance using standard rulers like horizon scale of sound at last scattering\cite{komatsu2011} and baryon acoustic oscillations\cite{Eisenstein2005}. Yet the determination of the expansion rate alone cannot give the crucial information which can differentiate between different dark energy models. That is to say, models with different physical origins but the same global expansion properties could not be separated.

 The other important probe is the cosmic growth which tests the evolution of the inhomogeneous part of the energy density. The growth of large scale structures, derived from the linear matter density contrast $ \delta(z) \equiv \delta\rho/\rho $ in the universe, serves as an important companion test. This can provide significant insight into the properties of dark energy which could possibly remove the degeneracy between various models.

The cosmic expansion history constrains the equation of state of dark energy. So, the usual practice is to parametrize the effective equation of state of the dark energy $w(z)\equiv\frac{p_{de}(z)}{\rho_{de}(z)}$  to mimic the  observational description of the expansion rate $H(z)$.  Similarly in the cosmic growth history, the matter density perturbation $ \delta(z)$ is constrained. The standard approach is to parametrize the growth function $f=\frac{dln\delta}{dln a}$ in terms of growth index $ \gamma $  to mimic the evolution of the inhomogeneous energy density. As we will describe in the next section, this kind of proposal of parametrizing $\delta$ in terms of $\gamma$ was first introduced by Peebles \cite{peb} and later very effectively expanded by Wang and Steinhardt \cite{ws:98} for dark energy models. Parametrizations  of $ \gamma $ have been done from different motivations\cite{many}.  In this article we introduce a new parametrization for growth index $\gamma$ motivated from Generalised  Chaplygin Gas (GCG).

GCG is a very interesting alternative to other proposals aiming to explain the observed accelerated expansion of the Universe\cite{gcg}. It is an exotic background fluid, characterised by the equation of state
\begin{equation} 
p_{gcg}=\frac{-A}{\rho_{gcg}^{\beta}},
\end{equation}
where $A$ and $\beta$ are constants. Within the framework of Friedmann-Robertson-Walker cosmology, this equation of state leads to the density evolution 
\begin{equation}
\rho_{gcg}=\rho_{gcg0}\left[A_{s}+(1-A_{s})(1+z)^{3(1+\beta)}\right]^{\frac{1}{1+\beta}},
\end{equation}
where $ A_{s} = \frac{A}{\rho_{gcg0}^{1+\beta}} $
with $\rho_{gcg0}$ being the present value of $\rho_{gcg}$. The choice of $A_{s}$ and $\beta$ uniquely specifies the GCG model. It is straightforward to check that $A_{s} = -w_{gcg}(z=0)$. The parameter $\beta$ plays an interesting role. For $(1+\beta) > 0$, at early times the energy density behaves as matter while at late times it behaves as a cosmological constant.  Whereas, for $(1+\beta) < 0$, GCG behaves as a cosmological constant to start with and in future it starts behaving like a matter fluid.  With this, the late time acceleration can be a transient phenomena. All these diverse possibilities make the GCG equation of state a unique one. 

The GCG model has been successfully confronted with various  obervational tests: high precision Cosmic Microwave Background Radiation data\cite{gcgcmb}, Supernova data\cite{gcgsn}, and gravitational lensing\cite{gcglens}. It was also shown that GCG can mimic dark energy models with phantom equation of state avoiding the pathology of violation of dominant energy condition\cite{gcgmnras}. Despite all these pleasing features, there are certain issues of  structure formation that plagues the dark matter sector of this model\cite{gcgdmprob}. Some alternate views are also suggested in that context\cite{gcgdmalt}. However, here we concentrate on the dark energy aspect of the model. In recent studies, it has been shown that GCG parametrization for dark energy equation of state consistently performs very well with different model selection criteria e.g $\chi^2$/dof, GoF, BIC, AIC etc, when compared to other parametrization for dark energy equation of state \cite{gcgcrit}.

 In this article we study the growth rate of the large scale structures in a cosmological scenario where GCG serves as the dark energy candidate. Following the  Wang and Steinhardt\cite{ws:98} ansatz,  we parametrize the growth index $ \gamma$ in terms of the GCG parameters $(A_{s},\beta)$.  To check the efficiency of our parametrization we compare the growth rate from two approaches (theoretical and parametrized) and obtain a very impressive agreement (less than 1\% difference). Subsequently, we use this new parametrization to portray the growth of structures in a wide range of dark energy models and show that this new parametrization can successfully mimic the growth index of a variety of dark energy models. We also constrain the dark energy behaviour using current data for growth mainly from galaxy redshift distortion and Lyman-$\alpha$ forest. We show that given our current knowledge of the present day density parameter $\Omega_{m0}$ from WMAP-7 measurements, growth data allow both transient as well as an eternally accelerating universe.
 
 The paper is planned as follows: The study of growth rate for GCG model, the formulation of the parametrization of  growth index and a comparison between the two approaches are discussed in Section 2. In  Section 3, the new parametrization is used to depict the growth rate for various dark energy models and a comparative study is done to see the difference between the proposed and the theoretical one. We present the compiled data set of currently available observations  and do a maximum likelihood analysis in Section 4. Finally, in Section 5, we conclude by summarising our results.

\section{Parametrization of the Growth Index}

The growth rate of large scale structures is derived from matter 
density perturbation $\delta=\delta\rho_m/\rho_m$ in the linear regime, 
satisfying  
\be\label{delta}
\ddot{\delta}+2\frac{\dot{a}}{a}\dot{\delta}-4\pi G\rho_{m}\delta=0 .\\
\ee
 While the equations that influence the background cosmology 
involving matter and GCG fluid are
\begin{eqnarray}\label{fldeqn}
\left(\frac{\dot{a}}{a}\right)^{2}=\frac{8\pi G}{3}(\rho_{m}+\rho_{gcg}),\\
2\frac{\ddot{a}}{a}+\left(\frac{\dot{a}}{a}\right)^{2}=-8\pi G\omega_{gcg}\rho_{gcg},
\end{eqnarray}
Here, the dot is derivative with respect to time `t' and $w_{gcg}$ is the equation of state for GCG given by, 
\be\label{wgcg}
w_{gcg}=\frac{-A_{s}}{\left[A_{s}+(1-A_{s})(1+z)^{3(1+\beta)}\right]}.
\ee
It is quite straightforward to change the variable in equation(\ref{delta}) 
from $t$ to $lna$ to obtain
\be\label{delta1}
(ln\delta)''+(ln\delta)'^{2}+ln\delta'\left[\frac{1}{2}-\frac{3}{2}w_{gcg}(1-\Omega_{m}(a))\right] = \frac{3}{2}\Omega_{m}(a),
\ee
where 
\be\label{om}
\Omega_{m}(a) = \frac{\rho_{m}}{\rho_{m}+\rho_{gcg}}.
\ee
With the help of energy conservation equation 
$d\rho = -3(\rho+p)dlna$ we further change the variable from
$lna$ to $\Omega_m(a)$ and recast the above equation in terms of the 
logarithmic growth factor $f \equiv \frac{dln\delta}{dlna}$ as 
\be\label{f}
3w_{gcg}\Omega_{m}(1-\Omega_{m})\frac{d~f}{d\Omega_{m}}+f^{2}+f\left[\frac{1}{2}-\frac{3}{2}w_{gcg}(1-\Omega_{m}(a))\right]  = \frac{3}{2}\Omega_{m}(a).
\ee

As we mention in the previous section, Wang and Steinhardt\cite{ws:98} had proposed an ansatz for the logarithmic growth factor:
\begin{equation}\label{fom}
f = \Omega_{m}(a)^{\gamma},
\end{equation}
 where $\gamma$ is termed as {\em the growth index parameter}. 
 This is an excellent fit to $ f $ for various cosmological models with specific choice of $ \gamma $.  For flat dark energy models with constant equation of state $w_0$, $\gamma$ is given by 
\begin{equation}\label{gw0}
\gamma=\frac{3(w_0-1)}{6w_0-5}.
\end{equation}
For $\Lambda$CDM reduces to $\frac{6}{11}$\cite{nesseris}. For matter 
dominated open universe model $\gamma=\frac{4}{7}$\cite{Fry,Lightman}. For dark energy models with 
slowly varying equation of state, the growth index parameter $ \gamma$ is found 
to be function of $ \Omega_m $ and $ w $. As the growth data spans 
over a range of redshift, people have parametrized 
$ \gamma $ as a function of redshift. In one such attempt
\cite{polarski}, $ \gamma $ was  proposed to be parametrized as 
$ \gamma(z)=\gamma_0+\gamma^{'} z$ (where 
$ \gamma^{'}\equiv\frac{d\gamma}{dz} (z=0)$)  to distinguish between
 models of dark energy and models of modified gravity. There are also other proposals from different motivations\cite{ishak,dossett}.

Here we parametrize $ \gamma $ in terms of GCG parameters $ A_s $ and
$\beta$. For that we assume the original ansatz by Wang and 
Steinhardt\cite{ws:98} 
\begin{equation}\label{fomgam}
f = \Omega_{m}^{\gamma(\Omega_{m})},
\end{equation}
where the growth index parameter $\gamma(\Omega_{m})$ can be Taylor 
expanded around $\Omega_{m} = 1$ as
\begin{equation}\label{gamtay}
 \gamma(\Omega_{m}) = \gamma|_{(\Omega_{m}=1)}+(\Omega_{m}-1)\frac{d\gamma}{d\Omega_{m}}|_{(\Omega_{m}=1)}+O(\Omega_{m}-1)^{2}.
\end{equation}
Now, equation (\ref{f}) can be rewritten in terms of $\gamma $ as
\begin{equation}\label{gamom}
3w_{gcg}\Omega_{m}(1-\Omega_{m})ln\Omega_{m}\frac{d\gamma}{d\Omega_{m}}-3w_{gcg}\Omega_{m}\left(\gamma-\frac{1}{2}\right)+\Omega_{m}^{\gamma}-\frac{3}{2}\Omega_{m}^{1-\gamma}+3w_{gcg}\gamma-\frac{3}{2}w_{gcg}+\frac{1}{2} = 0.
\end{equation}
Once we differentiate the above equation around $\Omega_{m} = 1$,
after some simple calculations, we find the zeroth order term in
 the expansion for $ \gamma $ as
 \begin{equation}\label{g0}
\gamma = 3\frac{(1-w_{gcg})}{(5-6w_{gcg})},
\end{equation}
which agrees with the case of dark energy models with constant
$ w_0 $ (equation(\ref{gw0})). Following the same technique, i.e, 
differentiating twice around $\Omega_{m} = 1$, we find the first order 
term in the expansion of $ \gamma $
\begin{equation}\label{g1}
\frac{d\gamma}{d\Omega_{m}}\mid_{(\Omega_{m}=1)} = \frac{3(1-w_{gcg})(1-\frac{3w_{gcg}}{2})}{125(1-\frac{6w_{gcg}}{5})^{3}}. 
\end{equation}
Substituting equations (2.13) and (2.14) in the expansion (\ref{gamtay}), $ \gamma $, till the first order, can be approximated as 
\begin{equation}\label{gamgcg}
 \gamma(\Omega_{m}) = 3\frac{(1-w_{gcg})}{(5-6w_{gcg})}+(1-\Omega_{m})\frac{3(1-w_{gcg})(1-\frac{3w_{gcg}}{2})}{125(1-\frac{6w_{gcg}}{5})^{3}}. 
\end{equation}
One can now put the expression for $ w_{gcg}$ ( given in equation (2.4)) and subsequently 
$ \gamma $ can be parametrized in terms of the GCG parameters $ A_s $, $ \beta $ and redshift $ z $. 

We define normalised growth function $g$ from the numerically obtained solution of equation(\ref{delta1}) as
\begin{equation}\label{g}
 g(z)\equiv \frac{\delta(z)}{\delta(0)}.
\end{equation}
Also the corresponding approximate normalised growth function from the parametrized form of $f$ (equation (\ref{fomgam})) is given by 
\begin{equation}\label{gzom}
 g_{th}(z) = \exp^{\int_1^{\frac{1}{1+z}}\Omega_{m}(a)^{\gamma}\frac{da}{a}},
\end{equation}
where $ \gamma $ is parametrized by equation(\ref{gamgcg}).

Ansatz (\ref{fomgam})along with (\ref{gamgcg}) provide an excellent approximation to the theoretical growth factor obtained by numerically solving equation(\ref{f}). To show this, we plot the difference between the numerically obtained solution of equation(\ref{f}) for the growth factor $f $ of GCG model and the approximated growth factor given by equation(\ref{fomgam}) and (\ref{gamgcg}) in the left panel of Figure 1. We also show the relative percentage difference in normalised growth function $g(z)$ (in the right panel) for various values of $A_s$ and $\beta$. The difference between the two approaches is quite impressive ($\leq 1\%$) as can be seen in the plots. 
\begin{figure}
\begin{center}
\begin{tabular}{|c|c|}
\hline
 & \\
{\includegraphics[width=2.6in,height=2.1in,angle=0]{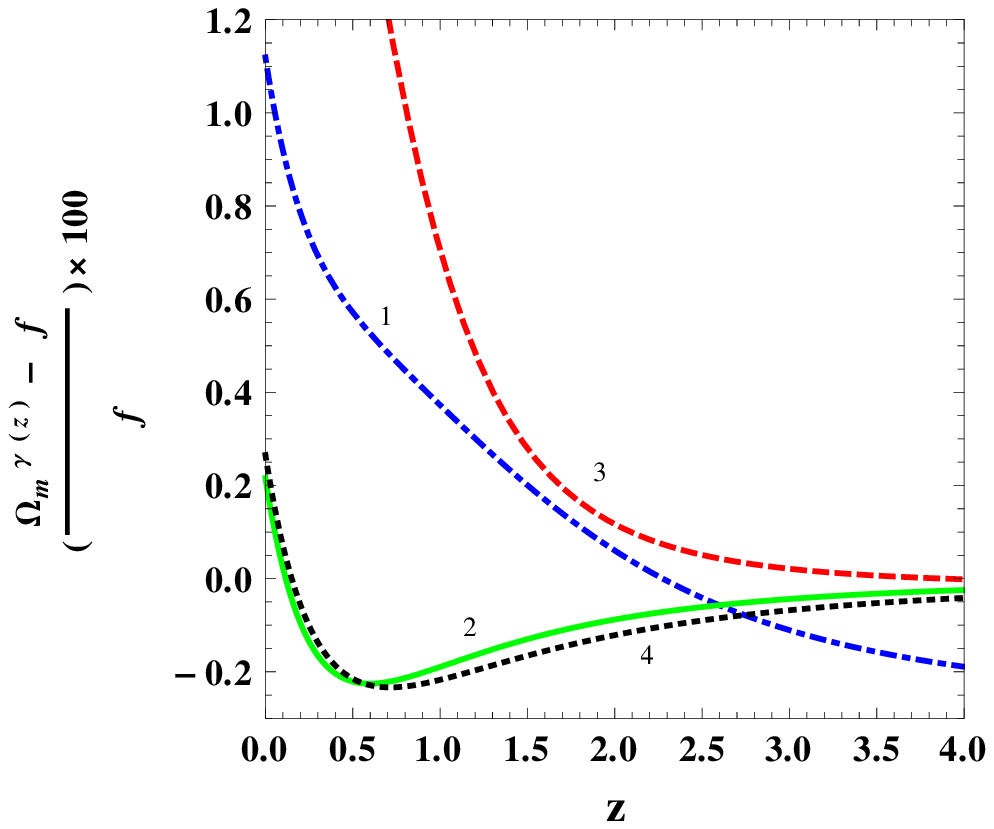}} &
{\includegraphics[width=2.6in,height=2.1in,angle=0]{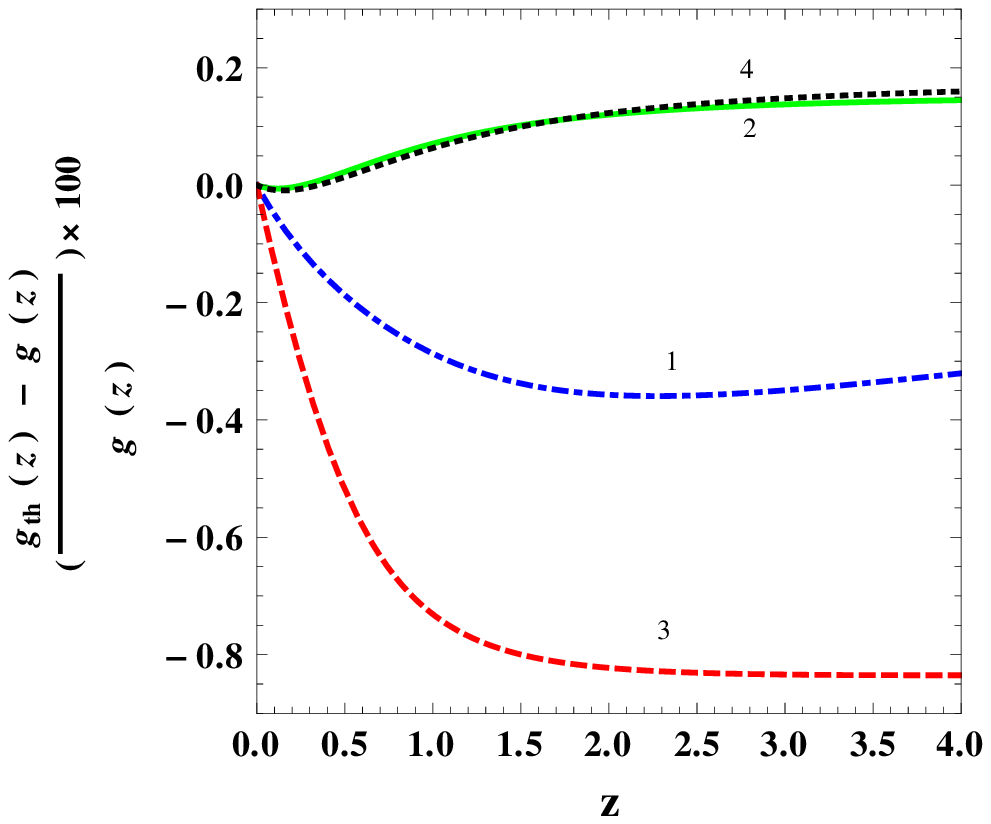}} \\
\hline
\end{tabular}
\caption{\label{fig1}
Plot of the relative percentage difference in theoretical (numerically calculated) and proposed growth rate as well as the normalised growth function.TOP-LEFT: In the left we have plotted $\frac{\Omega_m^{\gamma(z)}-f}{f}$ for various values of $A_s$ and $\beta$ in GCG model. The colour codes are: blue dotdashed (1) for $A_s=0.9$  $\beta= 0.1$, green-solid (2) for $A_s=0.9$ and $\beta= -1.05$, red-dashed (3) for $A_s=0.8$ and $\beta=0.6$ and black-dotted (4) for $A_s=0.8$ and $\beta=-1.02$. TOP-RIGHT: In the right we have plotted $\frac{g_{th}-g}{g}$ for the same values of $A_s$ and $\beta$ in GCG model. The colour codes used are same. 
}
\end{center}
\end{figure}

\section{Fitting the parametrized growth index to other dark energy models}

In the previous section we parametrize the growth index in terms 
of GCG parameters $ A_s $ and $ \beta $. In this section we use this parametrized form of $\gamma$ to find the growth rates for different dark energy models.
We match this parametrized growth rate with the one that is calculated directly from the model. 

\subsection{Dark Energy Models  with CPL parametrization}  
The basic equations governing the background cosmology for Dark Energy(DE) models are
\begin{eqnarray}
\left(\frac{\dot{a}}{a}\right)^{2}=\frac{8\pi G}{3}(\rho_{m}+\rho_{de}),\\
2\frac{\ddot{a}}{a}+\left(\frac{\dot{a}}{a}\right)^{2}=-8\pi G w_{de}\rho_{de},
\end{eqnarray}
where we have assumed $p_{de}=w_{de}\rho_{de}$. Each component of  energy density satisfies the conservation equation 
\begin{equation}
\frac{\dot{\rho_i}}{\rho_i} =-3H(1+\omega_i);~~~~  i~=~m,~de.
\end{equation}
Consequently, matter energy density scales as $a^{-3}$.  For dark energy we use 
the popular CPL parametrization\cite{Chevallier:2000, Linder:2002} for equation of state (EoS)
\begin{equation}
w_{de}(a)=w_0+w_a(1-a),
\end{equation} 
following which $\rho_{de}$ scales as $a^{-3(1+w_0+w_a)}e^{-3w_a(1-
a)}$. This conveniently includes the case of a constant EoS with 
$(w_0 = w, w_a = 0)$, and the $\Lambda$CDM model $(w_0 = -1, w_a = 0)$.
We find the linear density contrast $\delta$ 
from the equation(\ref{delta}) for such a background.   

Next, we fit the approximate growth factor $f$, calculated from the ansatz (\ref{fomgam}) and (\ref{gamgcg}), to the growth factor calculated for a particular dark energy model and find the fitting values for the GCG parameters $A_s$ and $\beta$. Thus a set of $(A_s, \beta)$ can represent the growth factor of a particular dark energy model.

\subsubsection{$\Lambda$CDM}

In the CPL parametrization, $\Lambda$CDM is represented by  
$w_0 = -1, w_a = 0$. The growth factor $f$ for $\Lambda$CDM is numerically calculated by solving equation(\ref{delta}) with this choice of $w_0$ and $w_a$.  Although the EoS of GCG reduces to cosmological constant for the choice $A_{s}=1$,  we fit the ansatz for $f$ given by equations (2.8) and (2.15) to the numerically obtained solution and find the fitting values for $A_{s}$ and $\beta$  to be $1.008$ and $-1.41$ respectively. 
\begin{figure}
\begin{center}
\begin{tabular}{|c|c|}
\hline
{\includegraphics[width=2.2in,height=1.5in,angle=0]{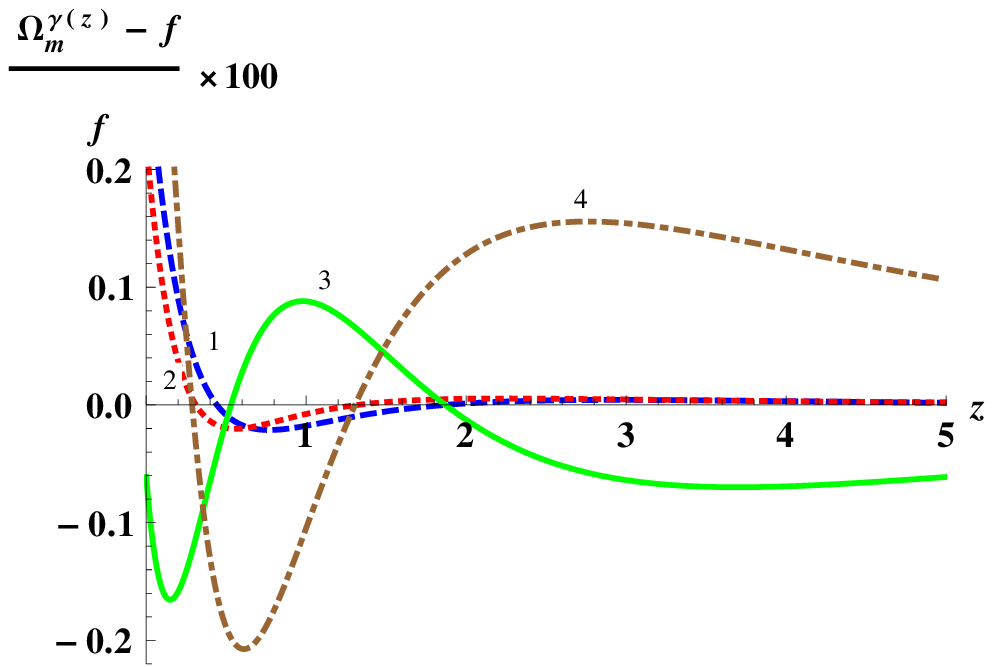}} &
{\includegraphics[width=2.2in,height=1.5in,angle=0]{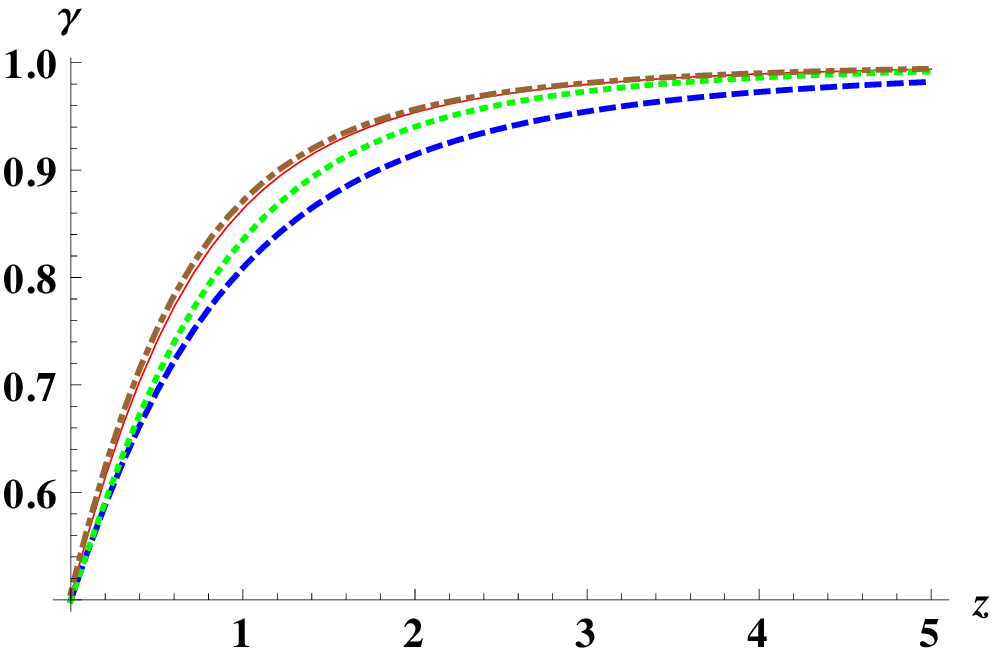}} \\
\hline
{\includegraphics[width=2.2in,height=1.5in,angle=0]{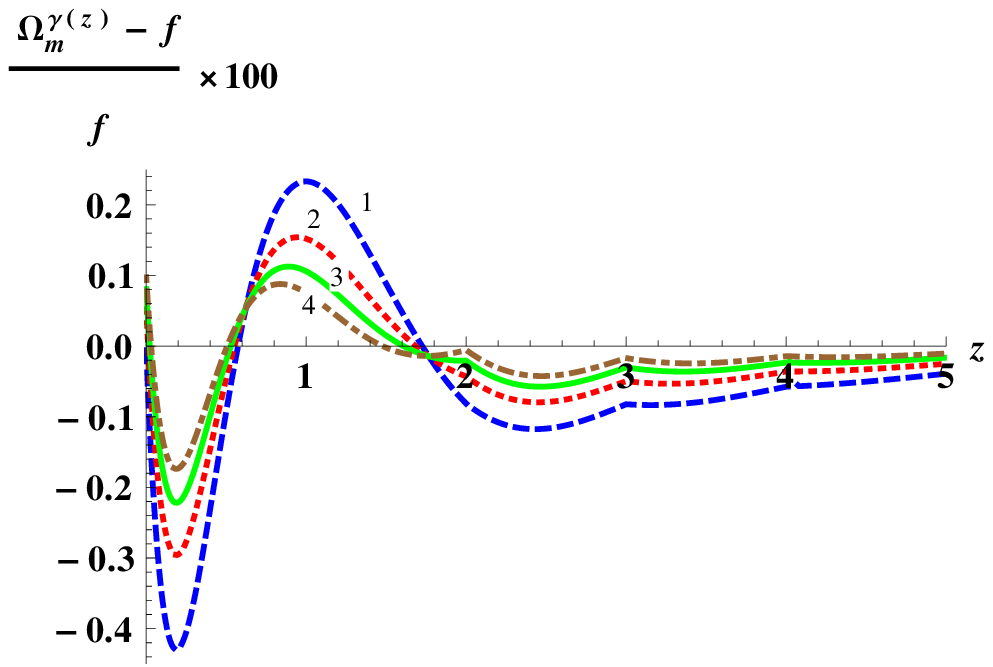}} &
{\includegraphics[width=2.2in,height=1.5in,angle=0]{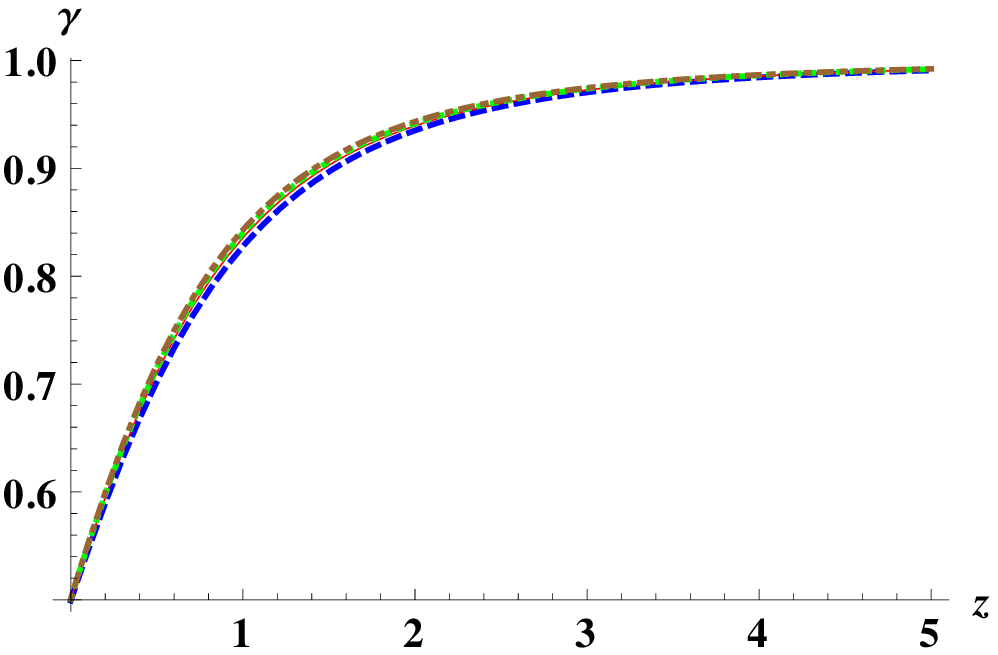}} \\
\hline
{\includegraphics[width=2.2in,height=1.5in,angle=0]{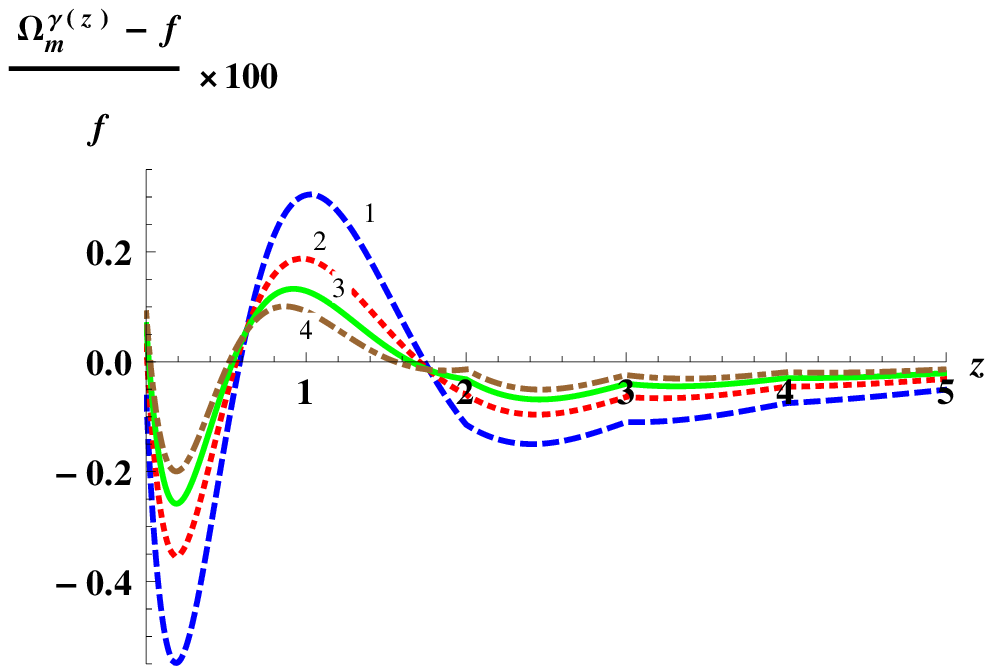}} &
{\includegraphics[width=2.2in,height=1.5in,angle=0]{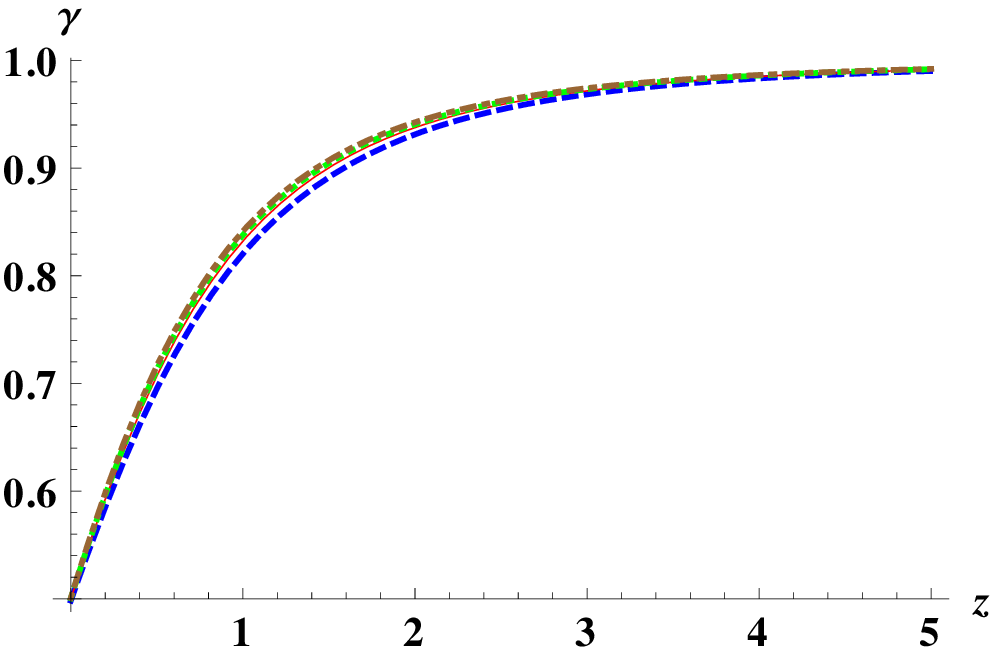}} \\
\hline
\end{tabular}
\caption{\label{fig2}
Theoretical fittings for GCG parametrization. In the left figures we have plotted the relative percentage difference $\frac{\Omega_m^{\gamma(z)}-f}{f}$ in order to compare the fit of the proposed parametrization to that of the growth factor $f$, that is numerically integrated from the growth ODE for various DE models.The figures on the right side are of the growth index for the corresponding models with the GCG parametrization. TOP-LEFT: Plot of the relative percentage difference against z for various QCDM models. The lines blue-dashed(1), red-dotted(2), green-solid(3) and brown-dotdashed(4) correspond to the quintessence models given in the Table 1 respectively from top to bottom.
TOP-RIGHT: $\gamma$ for various QCDM models for the best fit value of $\Omega_{m0}(0.2895)$.
MIDDLE-LEFT: Plot of the relative percentage difference against z for dark energy models with thawing scalar field following different potentials.  Here also the lines blue-dashed(1), red-dotted(2), green-solid(3) and brown-dotdashed(4) correspond to the thawing models given in the Table 1 respectively from top to bottom.
MIDDLE-RIGHT: $\gamma$ for Thawing scalar field models with different potentials for $\Omega_{m0}=0.2895$. 
BOTTOM-LEFT: For Dark energy models with tachyon field following different potentials  the relative percentage difference is plotted against z. We follow the same representation, i.e, the lines blue-dashed(1), red-dotted(2), green-solid(3) and brown-dotdashed(4) correspond to the tachyon models given in the Table 1 respectively from top to bottom. 
BOTTOM-RIGHT: $\gamma$ for Tachyon models following different potentials for the best fit value of $\Omega_{m0}$. 
}
\end{center}
\end{figure}

With this choice of $A_{s}$ and $\beta$ we find that the proposed parametrization is able to fit the theoretical value for $\Lambda$CDM to within $0.02\%$. The difference in growth rates between the two approaches is shown in Figure 2 (Red dotted line in the top left figure) and result of the fit is enlisted in Table 1. 

\subsubsection{Dark Energy Models with constant EoS and variable EoS}

Dark energy models with constant EoS are represented with CPL  parametrization as $w_0=w$
and $w_a=0$ while DE models with variable EoS would have nonzero 
values of both $w_0$ and $w_a$. Going through the same procedure as we have followed above, we solve numerically in equation (2.7) for theoretical $f$ for different DE models and fit that to the parametrized $f$. This gives us different fitting values for $(A_{s}, \beta)$ for different DE models, which we enlist in Table 1. With those fitting values, the GCG parametrization fits the theoretical values very accurately ( difference is $\leq 0.2\%$). The differences between the two $f$'s and the two $\gamma$'s for different DE models are shown in Figure 2 (top panel). The fitting values are enlisted in Table 1. 

\begin{table}
\begin{center}
\begin{tabular}{|c|c|c|}\hline
\multicolumn{3}{|c|}{\bfseries Parameters for various QCDM models.}\\ \hline
$\mathbf{(w_0,w_a)}$&$\mathbf{A_{s}}$&$\mathbf{\beta}$\\ \hline
$(-0.8,0)$&$0.806544$&$-0.991988$\\ \hline
$(-1,0)$&$1.0084$&$-1.41556$\\ \hline
$(-0.8,-0.3)$&$0.793615$&$-1.83504$\\ \hline
$(-1.1,0.22)$&$1.14599$&$-2.59154$\\ \hline \hline
\multicolumn{3}{|c|}{\bfseries Parameters for Thawing Scalar field models.}\\ \hline
$\mathbf{(\lambda,\Gamma)}$&$\mathbf{A_{s}}$&$\mathbf{\beta}$\\ \hline
$(1.0,0.0)$&$0.78782$&$-1.57618 $\\ \hline 
$(1.0,0.5)$&$ 0.83270$&$-1.55936 $\\ \hline
$(1.0,1.0)$&$0.85769 $&$-1.55919 $\\ \hline
$(1.0,1.5)$&$ 0.87423$&$-1.56673 $\\\hline\hline
\multicolumn{3}{|c|}{\bfseries Parameters For Tachyon models}\\ \hline 
$\mathbf{(\lambda_t,\Gamma_t)}$&$\mathbf{A_{s}}$&$\mathbf{\beta}$\\ \hline
$(1.0,0.0)$&$0.747729 $&$ -1.59938$\\ \hline
$(1.0,0.5)$&$0.813433 $&$-1.56497 $\\ \hline
$(1.0,1.0)$&$ 0.845355$&$-1.55841 $\\ \hline
$(1.0,1.5)$&$0.865417 $&$ -1.56237$\\ \hline\hline
\end{tabular}
\caption{\label{table:Expfit}
Values for parameters $A_{s}$ and $\beta$ in GCG parametrization that fit the actual growth for quintessence dark energy models (QCDM) with constant and variable equation of state, thawing scalar field models with different potentials, and tachyon models with different potentials.
}
\end{center}
\end{table}
\subsection{Thawing Scalar field models}
Now we test the effectiveness of our parametrization for models where slow rolling scalar field or thawing scalar field plays the role of dark energy. In this model,  at early times the scalar field is nearly frozen at $w = -1$ due to the large Hubble damping. Its energy density is nearly constant at this  stage and its contribution to the total energy density of the universe is also nearly negligible. But as radiation/matter rapidly  dilutes due to expansion of the universe, the background energy density becomes comparable to the scalar field energy density and the field breaks away from its frozen state evolving slowly to the region with larger values of equation of state parameter. 

The dynamical equation for the scalar field is
\begin{equation}\label{scalar}
\ddot{\phi}+3H\dot{\phi}+\frac{dV}{d\phi}=0,
\end{equation}
where V is the potential for the scalar field and H is the Hubble parameter described by the  background cosmology as
\begin{equation}\label{bc}
3H^2=\rho_m+\rho_\phi.
\end{equation}
These equations describing the background dynamics can be expressed as 
a set of autonomous equations\cite{sen}:
\begin{eqnarray}
\label{gammaprime}
\gamma_\phi^\prime &=& -3\gamma_\phi(2-\gamma_\phi) + \lambda(2-\gamma_\phi)\sqrt{3 \gamma_\phi
\Omega_\phi},\\
\label{Omegaprime}
\Omega_\phi^\prime &=& 3(1-\gamma_\phi)\Omega_\phi(1-\Omega_\phi),\\
\label{lambda}
\lambda_\phi^\prime&=&-\sqrt{3}\lambda_\phi^2(\Gamma-1)\sqrt{\gamma_\phi
\Omega_\phi}.
\end{eqnarray}
(prime denotes the derivative with respect to $ln a$)\\
where $\Omega_\phi$,  $ \gamma_\phi$, $ \lambda_\phi $ and $ \Gamma $ are defined as follows:
\begin{eqnarray}
\Omega_\phi = \frac{\rho_{\phi}}{3H^2}=\frac{\phi^{\prime 2}}{6} + \frac{V(\phi)}{3H^2},
~~~~~\gamma_\phi = (1+w_{\phi})= \frac{\phi^{\prime 2}}{3\Omega_\phi},\\
\lambda = -\frac{1}{V}\frac{dV}{d\phi},
~~~~\Gamma = V \frac{d^2 V}{d\phi^2}/\left(\frac{dV}{d\phi} \right)^2.
\end{eqnarray}

It is quite straight forward to express the Hubble parameter 
(equation(\ref{bc})) in terms of $\Omega_\phi$ as 
\begin{equation}
h^2(a)=\frac{H^2(a)}{H_0^2}=\frac{1-\Omega_{\phi 0}}{1-\Omega_{\phi}} a^{-3},
\label{hz}
\end{equation}
Hence, once $\Omega_\phi$ is determined from the above set of 
equations for various potential $V$, we can easily find the behaviour of the Hubble parameter and thereby the evolution of the growth factor $f$ for the thawing scalar field. We consider various types of potentials e.g $V =\phi$, $V = \phi^2$, $V = e^\phi$ and $V = \phi^{-2}$ , characterized by $\Gamma = 0$, $\frac{1}{2}$ , $1$ and $\frac{3}{2}$ respectively to obtain various growth rates for the thawing model.

To fit the parametrization we follow the same 
technique adopted for the dark energy models and consequently find a set of fitting values for $ (A_{s},\beta) $ corresponding to various potentials mentioned above. The results of the fits are provided in Figure 2 (middle panel) and Table 1.  Our parametrization fits thawing scalar field models with different potentials convincingly ( with more than $99\%$ accuracy).

\subsection{Thawing Tachyon models}
Our next target is Thawing tachyon models i.e, we test our proposal against these types of model.
The tachyon field is specified by the Dirac-Born-Infeld (DBI) type
of action \cite{sen}

\begin{equation}
{\mathcal{S}}=\int {-V(\phi)\sqrt{1-\partial^\mu\phi\partial_\mu\phi}}\sqrt{-g} d^4x.
\label{Taction1}
\end{equation}
The equation of motion which follows from (\ref{Taction1}) is
\begin{equation}
\ddot{\phi}+3H\dot{\phi}(1-\dot{\phi}^2)+\frac{V'}{V}(1-\dot{\phi}^2)=0,
\end{equation}
where $H$ is the Hubble parameter. In a similar fashion like the 
thawing scalar field model, the evolution equations for the tachyon 
can also be cast in a set of autonomous equations\cite{sen}
\begin{eqnarray}
\gamma_t'=-6\gamma_t(1-\gamma_t)+2\sqrt{3\gamma_t\Omega_t}\lambda_t
(1-\gamma_t)^\frac{5}{4},
\label{tachgamma}\\
\Omega_t'=3\Omega_t(1-\gamma_t)(1-\Omega_t),
\label{tachomega}\\
\lambda_t'=-\sqrt{3\gamma_t\Omega_t}\lambda_t^2(1-\gamma_t)^\frac{1}{4}(\Gamma-\frac{3}{2}).
\label{tachlambda}
\end{eqnarray}
(prime denotes the derivative with respect to $ln a$)\\
Here, the density parameter of the tachyon field $\Omega_t$, the EoS 
of the tachyon field $ \gamma_t \equiv 1+w_t$,  $ \lambda_t $ and 
$ \Gamma $ are defined as follows
\begin{eqnarray}
\Omega_t = \frac{\frac{V}{3H^2}}{\sqrt{1+\dot{\phi}^2}} ,
~~~~~\gamma_t =\dot{\phi}^2 ,\\
\lambda_t = -\frac{1}{V^{3/2}}\frac{dV}{d\phi},
~~~~\Gamma = V \frac{d^2 V}{d\phi^2}/\left(\frac{dV}{d\phi} \right)^2.
\end{eqnarray}
As done in the thawing scalar field case once $\Omega_t $ is solved from the above set of autonomous equations it is 
 easy  to find the behaviour of the Hubble parameter 
for various potentials. Subsequently, the evolution of the growth 
factor for the tachyon field rolling down different potentials could be 
solved. We consider similar potentials for tachyon field as 
well, i.e, $V =\phi$, $V = \phi^2$, $V = e^\phi$ and $V = \phi^{-2}$ , 
characterized by $\Gamma = 0, \frac{1}{2} , 1$ and $\frac{3}{2}$ 
respectively.

For fitting the parametrization we follow the same path and find the set of $ (A_{s},\beta) $ corresponding to various potentials.  Like the other cases, here also we find that our parametrization fits the theoretical value to within $0.55\%$.  We enlist the results in Figure 2 (bottom panel) and table 1.

We fit the growth function of various dark energy models with our fitting function for growth index given in equation (2.15). In figure 2, we show the difference between our fitting function and the actual growth index for a given dark energy model. The figure shows that difference is extremely small for redshifts as large as $z=5$ which is sufficiently high for observational data from the growth history of the universe. In other words, our fitting function for growth index    is quite accurate for redshifts that can be probed by present or future growth data.

We also study the sensitivity of the fitting procedure on cosmological parameter e.g $\Omega_{m0}$. In Figure 3, we show this for GCG case as well as for the CPL parametrization by taking different values for $\Omega_{m0}$. The figure shows that the fitting does not depend on the choice of $\Omega_{m0}$.
\begin{figure}
\begin{center}
\begin{tabular}{|c|c|}
\hline
 & \\
{\includegraphics[width=2.6in,height=2in,angle=0]{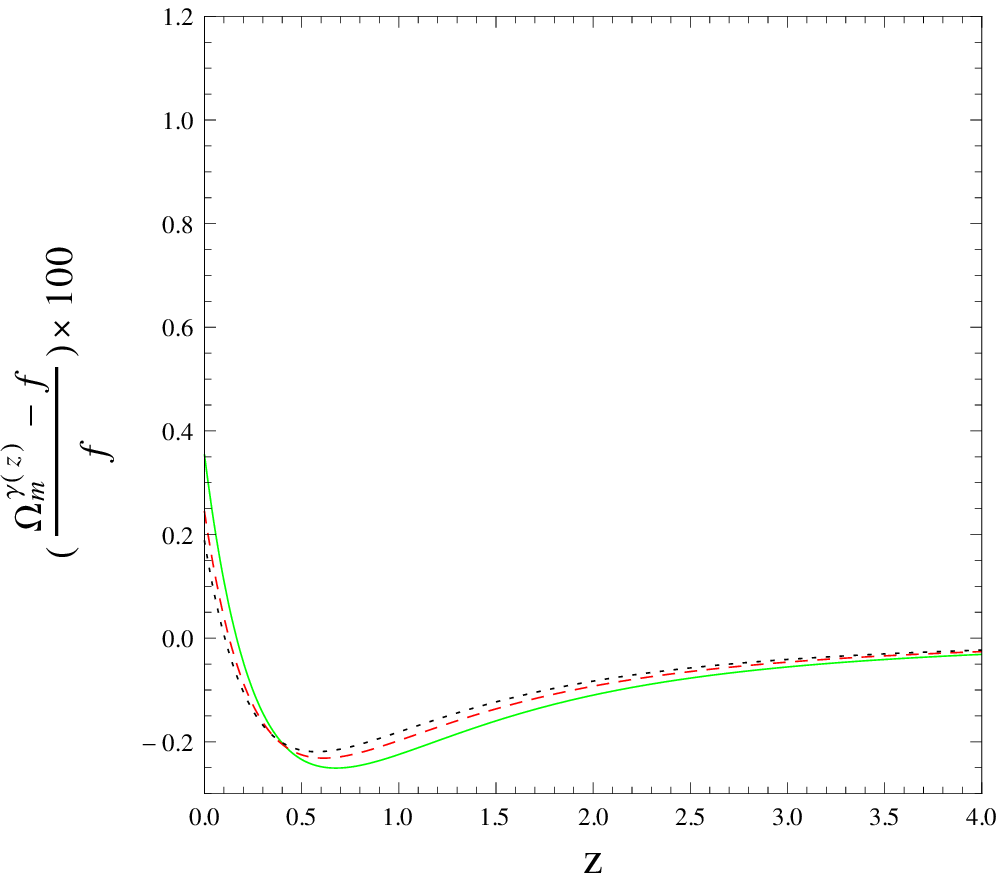}}&
{\includegraphics[width=2.6in,height=2in,angle=0]{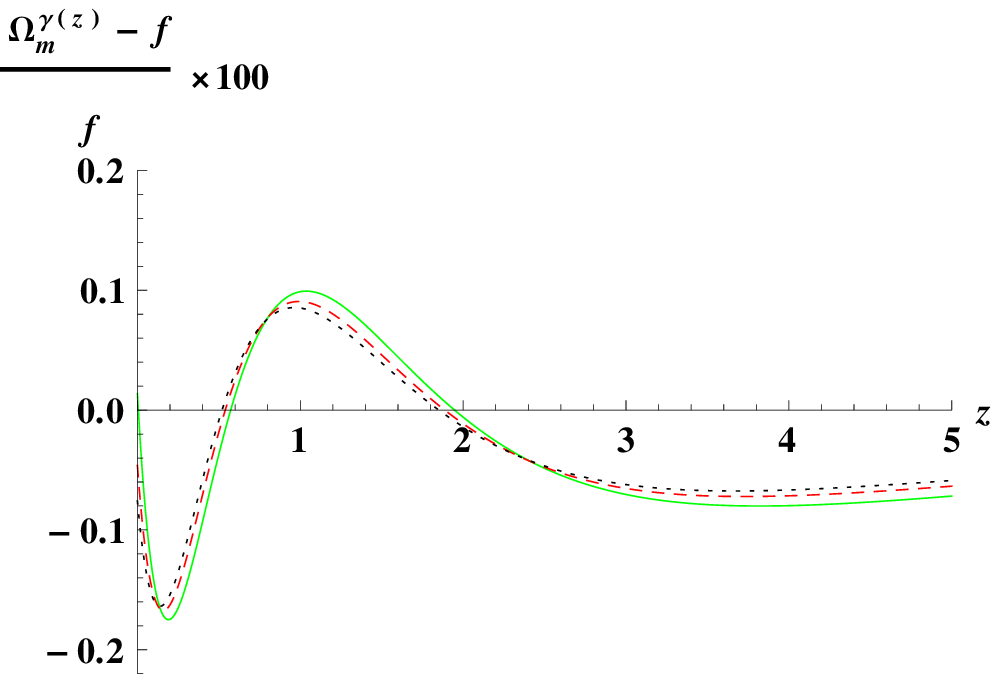}}
\\
\hline
\end{tabular}
\caption{Plot of the relative percentage difference in theoretical (numerically calculated) and proposed growth rate. LEFT: In the left we have plotted $\frac{\Omega_m^{\gamma(z)}-f}{f}$ for GCG model with $A_s=0.9$ and $\beta=-1.05$  for different values of $\Omega_{m0}$. The color codes are green(solid), red(dashed) and black(dotted) for $\Omega_{m0}$ = 0.25, 0.28 and 0.3 respectively. RIGHT: In the right we have plotted $\frac{\Omega_m^{\gamma(z)}-f}{f}$ for CPL model with $w_0= -0.8$ and $w_a=-0.3$ for different values of $\Omega_{m0}$. The color codes used are same.}
\end{center}
\end{figure}

\section{Observational Constraints}

In this section we constrain our model  with the current observational data . We perform a maximum likelihood analysis to find the best fit parameters for $ \gamma(A_{s},\beta,\Omega_{m0})$.  The growth data is given in Table 2, along with the references.  The data in Table 2, obtained 
from references \cite{Tegmark:2006az},\cite{Ross:2006me},\cite{daAngela:2006mf},\cite{bla2011} involve 
redshift distortion parameter $\beta$, which is related to the growth 
rate as $\beta=f/b$ (b is bias that measures how galaxy traces mass 
density field). The data for parameters $\beta$ and $b$ at various 
redshifts are given in references \cite{Di Porto:2007},\cite{bla2011}, where 
similar analyses are performed using different parametrizations. A 
catch in using this data to constrain other cosmological models is that 
at various steps in the process of analysing or converting the data 
$\Lambda$CDM model is used. So, if one uses the data to constrain 
other models, in particular the modified gravity models like DGP, one 
has to redo all steps assuming that model, starting from original 
observation. The data from reference \cite{McDonald:2004xn} does not involve $\beta$, 
but is obtained from various power spectrum amplitudes of Ly-$\alpha$ 
forest data in SDSS. 
We thus define 
\be \chi_f^2
(\Omega_{0m},\gamma) = \sum_i
\left[\frac{f_{obs}(z_i)-f_{th}(z_i,\gamma)}{\sigma_{f_{obs}}}\right]^2
\label{chif2} \ee 
where $f_{obs}$ and $\sigma_{fobs}$ are obtained
from Table 2 while $f_{th}(z_i,\gamma)$ is obtained from equations 
(\ref{fomgam}) and (\ref{gamgcg}).
\begin{table}[h!]
\begin{center}
\caption{Currently available data for linear growth rate $f_{obs}$ used in our analysis. $z$ is redshift; $\sigma$ is the 1$\sigma$ uncertainty of the growth rate data. \label{fdata}}
\vspace{6pt}
\begin{tabular}{c c c c c}
\hline\hline
$z$ & $f_{obs}$ & $\sigma$ & Ref. \\
\hline
0.15 & 0.51 & 0.11 & \cite{Hawkins:2002sg},\cite{Verde:2001sf} \\
0.22 & 0.60 & 0.10 & \cite{bla2011} \\
0.32 & 0.654 & 0.18 & \cite{reyes1003.2185}\\
0.35 & 0.70 & 0.18 & \cite{Tegmark:2006az} \\
0.41 & 0.70 & 0.07 & \cite{bla2011} \\
0.55 & 0.75 & 0.18 & \cite{Ross:2006me} \\
0.60 & 0.73 & 0.07 & \cite{bla2011} \\
0.77 & 0.91 & 0.36 & \cite{guz2008} \\
0.78 & 0.70 & 0.08 & \cite{bla2011} \\
1.4 & 0.90 & 0.24 & \cite{daAngela:2006mf} \\
3.0 & 1.46 & 0.29 & \cite{McDonald:2004xn}\\
\hline\hline
\end{tabular}
\end{center}
\end{table}

Another dependable indirect observational probe for the density contrast
$\delta(z)$ is the redshift dependence of the rms mass fluctuation
$\sigma_8 (z)$. It is defined as 
\be\label{sigrz} 
\sigma^2(R,z)=\int_0^\infty W^2(kR)\Delta^2(k,z)\dfrac{dk}{k},
\ee where 
\begin{eqnarray}
W(kR)&=&3\left(\frac{\sin (kR)}{(kR)^3}-\frac{\cos
(kR)}{(kR)^2}\right) r \label{wkr}, \\ 
\Delta^2 (kz)&=&4\pi k^3 P_\delta (k,z),
\label{delkz} 
\end{eqnarray}
with $R=8h^{-1}Mpc$ and $P_\delta(k,z)$ the mass power spectrum at redshift $z$. The function $\sigma_8 (z)$ is related to $\delta (z)$ as 
\be
\sigma_8(z)=\frac{\delta(z)}{\delta(0)} \sigma_8(z=0),
\label{s8del} 
\ee 
which leads to
\be s_{th}(z_1,z_2)\equiv\frac{\sigma_8(z_1)}{\sigma_8(z_2)}=\frac{\delta(z_1)}{\delta(z_2)}=\frac{e^{\int_1^{\frac{1}{1+z_1}}\Omega_m^\gamma
(a)\frac{da}{a}}} {e^{\int_1^{\frac{1}{1+z_2}}\Omega_m^\gamma
(a)\frac{da}{a}}}\label{sthrats8}, \ee
where we have used equation(\ref{gzom}).  From the redshift evolution of the flux power spectra for Ly-$\alpha$ forest, one gets the measurements for $\sigma_{8}(z)$
\cite{Viel:2004bf,Viel:2005ha,vimos}. These data points are given
in Table 3 along with the corresponding reference sources.

\begin{table}[!t]
\begin{center}
\caption{The currently available data for the rms fluctuation
$\sigma_8 (z)$ at various redshifts and references. Notice that
the data from Ref. \cite{vimos} were obtained using the normalized
by $a$ growth factor $\delta$ and in our analysis we took this
into account. \label{table2}}
\begin{tabular}{cccc}
\hline \hline\\
\hspace{6pt}   $z$  &\hspace{6pt}  $\sigma_8 $ & $\sigma_{\sigma_8 }$ &\hspace{6pt}  \textbf{Ref.} \\
\hline\\
\hspace{6pt}   2.125 &\hspace{6pt}  0.95 &\hspace{6pt}  0.17 & \cite{Viel:2004bf}  \\
\hspace{6pt}   2.72 &\hspace{6pt}  0.92 &\hspace{6pt}  0.17 &   \\
\hline\\
\hspace{6pt}   2.2  &\hspace{6pt}  0.92 &\hspace{6pt}  0.16 & \cite{Viel:2005ha}  \\
\hspace{6pt}   2.4  &\hspace{6pt}  0.89 &\hspace{6pt}  0.11 &   \\
\hspace{6pt}   2.6  &\hspace{6pt}  0.98 &\hspace{6pt}  0.13 &   \\
\hspace{6pt}   2.8  &\hspace{6pt}  1.02 &\hspace{6pt}  0.09 &   \\
\hspace{6pt}   3.0  &\hspace{6pt}  0.94 &\hspace{6pt}  0.08 &   \\
\hspace{6pt}   3.2  &\hspace{6pt}  0.88 &\hspace{6pt}  0.09 &   \\
\hspace{6pt}   3.4  &\hspace{6pt}  0.87 &\hspace{6pt}  0.12 &   \\
\hspace{6pt}   3.6  &\hspace{6pt}  0.95 &\hspace{6pt}  0.16 &   \\
\hspace{6pt}   3.8  &\hspace{6pt}  0.90 &\hspace{6pt}  0.17 &   \\
\hline\\
\hspace{6pt}   0.35 &\hspace{6pt}  0.55 &\hspace{6pt}  0.10 & \cite{vimos}  \\
\hspace{6pt}   0.6  &\hspace{6pt}  0.62 &\hspace{6pt}  0.12 &   \\
\hspace{6pt}   0.8  &\hspace{6pt}  0.71 &\hspace{6pt}  0.11 &   \\
\hspace{6pt}   1.0  &\hspace{6pt}  0.69 &\hspace{6pt}  0.14 &   \\
\hspace{6pt}   1.2  &\hspace{6pt}  0.75 &\hspace{6pt}  0.14 &   \\
\hspace{6pt}   1.65 &\hspace{6pt}  0.92 &\hspace{6pt}  0.20 &   \\
\hline \hline \\
\end{tabular}
\end{center}
\end{table}

Using the data of Table 3 we calculate the corresponding $\chi_s^2$
defined as 
\be \chi_s^2 (\Omega_{0m},\gamma)=\sum_i
\left[\frac{s_{obs}(z_i,z_{i+1})-s_{th}(z_i,z_{i+1})}{\sigma_{s_{obs,i}}}\right]^2
\label{chi2s} \ee 
where $\sigma_{s_{obs,i}}$ is derived by error
propagation from the corresponding $1\sigma$ errors of $\sigma_8
(z_i)$ and $\sigma_8 (z_{i+1})$ while $s_{th}(z_i,z_{i+1})$ is
defined in equation (\ref{sthrats8}) (We follow the method prescribed in \cite{nesseris}). The combined
$\chi_{tot}^2 (\Omega_{0m},\gamma)$ as 
\be 
\chi_{tot}^2(\Omega_{0m},\gamma)\equiv
\chi_{f}^2(\Omega_{0m},\gamma)+\chi_{s}^2(\Omega_{0m},\gamma) \label{chitot2} 
\ee

Minimising $\chi_{tot}^2$ with respect to all the parameters $A_{s}, \beta$ and $\Omega_{m0}$, we found the best fit growth rate $f(z)$. The corresponding best fit values of the parameters 
are $A_{s}=0.764$, $\beta=-1.436$ and $\Omega_{m0}=0.2895$. In Figure 4 we have presented the best fit for $f$ with corresponding $1\sigma$ errors 
(shaded region) along with the cosmological growth data of Table 2. In the same plot we have also presented all the models so far considered in this article with their corresponding choice of $(A_{s},\beta)$ given 
in Table 1. The intention is to find the models permissible within $1\sigma$ confidence level of the available data. 
Clearly, the figure shows that all the models including $\Lambda$CDM are well within the $1\sigma$ region. Only one model with CPL parametrization is barely crossing the $1\sigma$ region 
(green line)  at low redshifts. 

We further obtain the constraints on our three model parameters, e.g, $A_{s}$, $\beta$ and $\Omega_{m0}$. First we marginalise over the parameter $\Omega_{m0}$ assuming the  WMAP7 bound \cite{komatsu2011}  on $\Omega_{m0}$, ($\Omega_{m0}=0.2669\pm0.0288$), and get the best fit values for $A_{s}$ and  $\beta$ as 0.891 and -1.191 respectively. This can be seen from the contours presented in right panel of Fig.4.

\begin{figure}
\begin{center}
\begin{tabular}{|c|c|}
\hline
 & \\
{\includegraphics[width=2.6in,height=2in,angle=0]{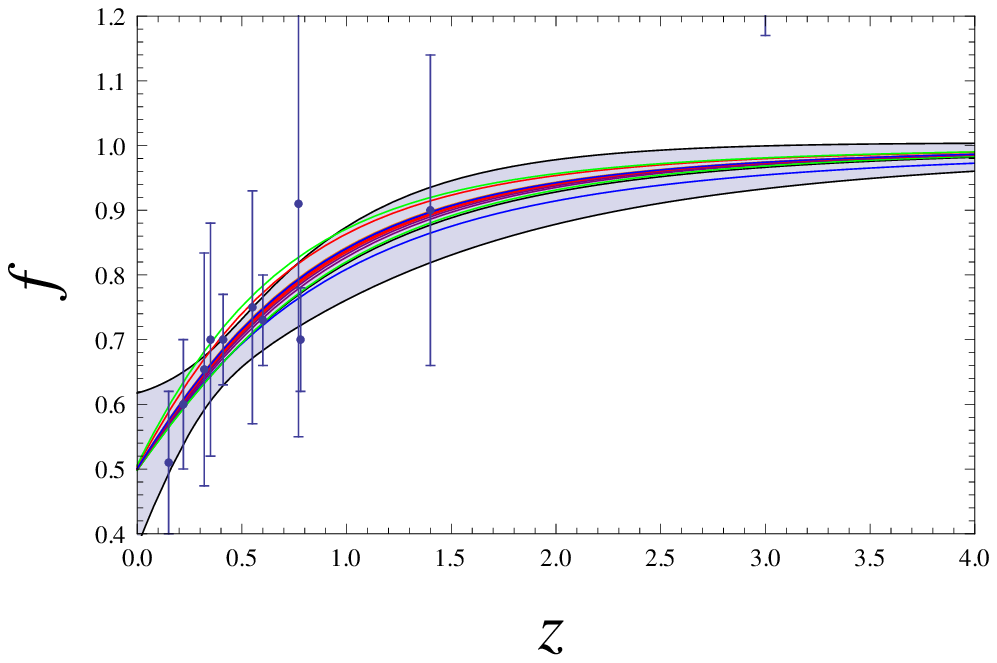}}&
{\includegraphics[width=2.6in,height=2in,angle=0]{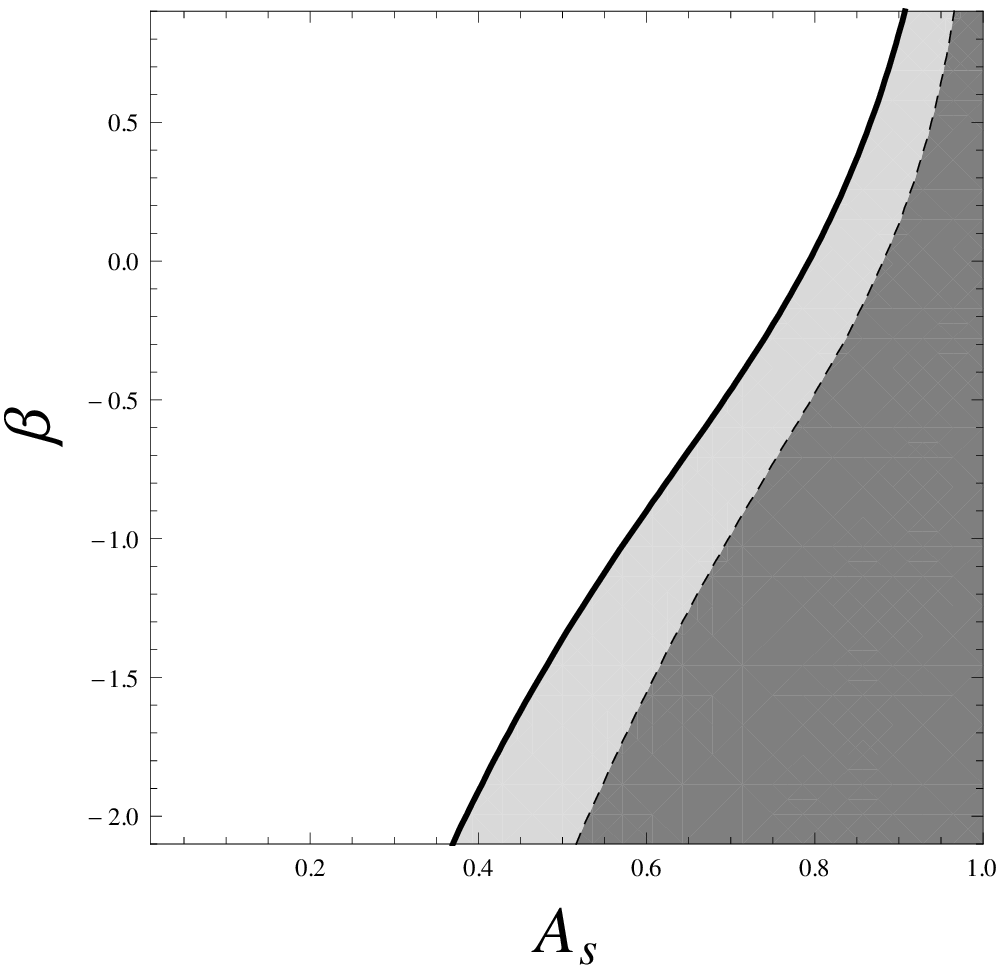}}
\\
\hline
\end{tabular}
\caption{LEFT: The cosmological data for the growth rate
$f(z)$ along with the best fit $f=\Omega_m(z)^\gamma$
and the corresponding $1\sigma$ errors (shaded region). 
All the models including $\Lambda$CDM
are well within the $1\sigma$ region. Only one model with CPL parametrization has barely crossed the $1\sigma$ region 
(green line)  at low
redshifts. RIGHT: Contour Plot for $(A_{s},\beta)$ marginalising $\chi_{tot}^2$ over $\Omega_{m0}$. The dark grey area  represents the allowed region at $1\sigma$ confidence level while the region bounded by the solid line represents that at $2\sigma$ confidence level. } \label{fig3}
\end{center}
\end{figure}

\begin{figure}
\begin{center}
\begin{tabular}{|c|c|}
\hline
 & \\
{\includegraphics[width=2.6in,height=2in,angle=0]{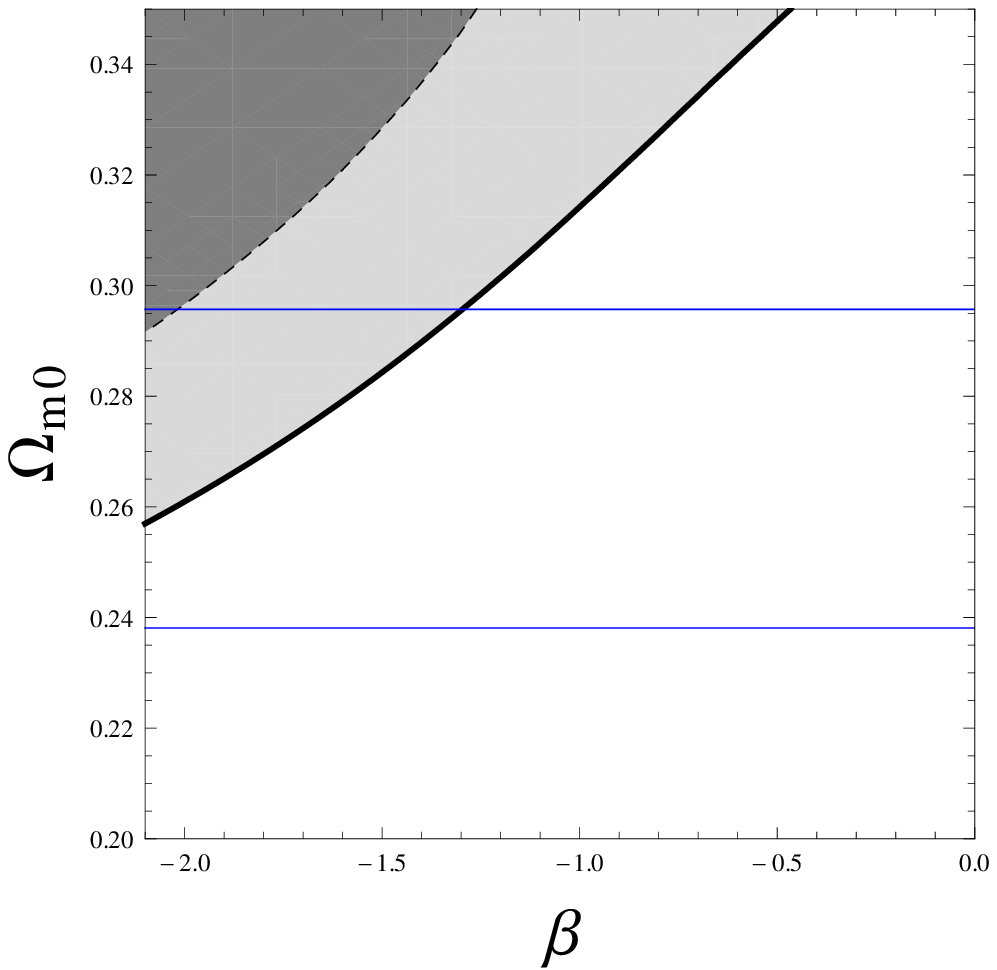}} &
{\includegraphics[width=2.6in,height=2in,angle=0]{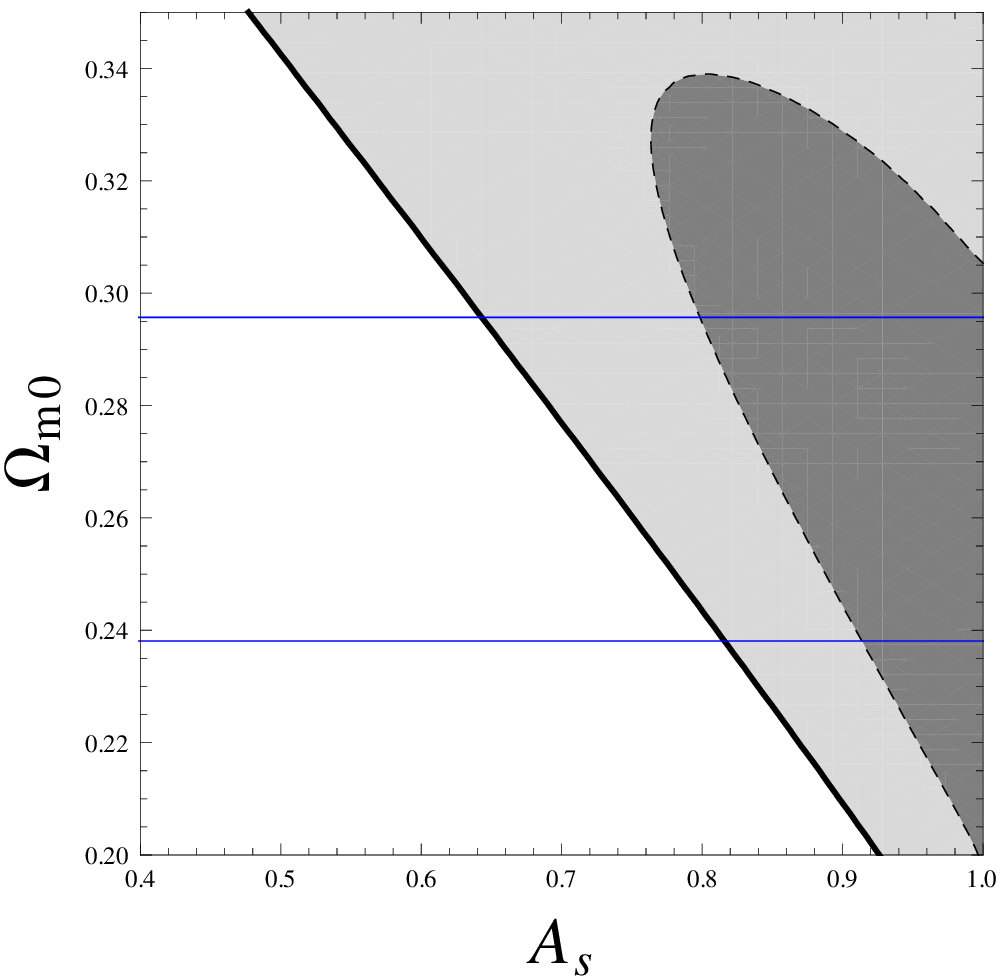}} \\
\hline
\end{tabular}
\caption{\label{fig:mchi}
Contour Plot for marginalised $\chi_{tot}^2$ with the bound on $\Omega_{m0}$ from 7 year WMAP data. LEFT: Contour Plot for $(\beta,\Omega_{m0})$ marginalising $\chi_{tot}^2$ over $A_{s}$.
RIGHT:Contour Plot for $(A_{s},\Omega_{m0})$ marginalising $\chi_{tot}^2$ over $\beta$. The dark grey area  represents the allowed region at $1\sigma$ confidence level while the region bounded by the solid line represents that at $2\sigma$ confidence level. }
\end{center}
\end{figure}

We further marginalise $\chi_{tot}^2$ over $A_{s}$ and $\beta$ respectively. In figure 5 we show the corresponding contour plots. We marginalise $\chi_{tot}^2$ over $A_{s}$ and plot the contours in $(\beta,\Omega_{m0})$ plane. This along with the bound from 7 year WMAP data\cite{komatsu2011} on $\Omega_{m0}$ results an upper limit of $\beta$,  $\beta \leq -1.3$ at $2\sigma$ confidence level. As we mentioned earlier $(1+\beta)< 0$ represents cosmological evolution with transient late time acceleration, hence this result shows that growth history of the universe may allow a transient accelerating universe. This is an interesting outcome from our analysis. 

Marginalising $\chi_{tot}^2$ over $\beta$, the contours in $(A_{s},\Omega_{m0})$ plane along with the bound from 7 year WMAP data not only allows the $\Lambda$CDM ($A_{s}=1$) but also gives a lower bound on $A_{s}$,   $A_{s} \geq 0.62$ at the $2\sigma$ confidence level.

\section{Conclusion}
It has been realised over the last few years that
observations probing cosmic expansion alone is insufficient to identify any particular theory explaining the late time acceleration. The study of the growth of large scale structures has emerged as a complementary probe for differentiating between the different dark energy models. In this work we explore this aspect using GCG as a possible parametrization for dark energy equation of state. We study the growth of the matter perturbation in GCG model and then propose a parametrization of the growth index $\gamma$ in terms of the GCG parameters that would mimic the original growth history of a GCG model. The new $\gamma$ is constructed following the same ansatz as prescribed by Wang-Steinhardt \cite{ws:98}. We test the proposed parametrization against other candidates for dark energy. We find that GCG parametrization can fit the growth evolutions of different dark energy models extremely well. 

We constrain our model with latest data from redshift distortion of 
galaxy power spectra and the rms mass fluctuation ($\sigma_8$) from Ly-$\alpha$ surveys. Our compiled set of 11 data points in the growth data set (Table 2) includes the 4 latest Wiggle-Z survey data\cite{bla2011}. In the other set of 17 data points, the growth rate is derived from $\sigma_8$ data from the power spectrum of Ly-$\alpha$ surveys. Maximum likelihood analysis with all these 28 data for the GCG parametrization shows  that the growth in various DE models represented by GCG parametrization falls within $1\sigma$ allowed region. Hence, with the current error bars for growth measurements, it is not possible to distinguish different dark energy models. Moreover, marginalised $\chi^2$ analysis puts an upper and lower bound on the GCG parameters $\beta$ and $A_{s}$ which are respectively $-1.3$ and $0.62$ within $2\sigma$ confidence level.
From this analysis, one concludes that the growth measurements allow both the transient late time acceleration as well as the concordance $\Lambda$CDM model. As we mentioned earlier, while minimising $\chi^{2}_{tot}$ for all the three parameters ($A_{s}, \beta$ and $\Omega_{m0}$) as well as minimising marginalised $\chi^{2}_{tot}$ over $\Omega_{m0}$ and $A_{s}$ respectively, the best fit values for $\beta$ are coming out to be less than -1.  We also minimize the $\chi^2$ over $A_{s}, \beta$ and $\Omega_{m0}$ in the region $1+\beta >0$. The $\chi^2_{min}$ in this case is 6.311. But when we allow $1+\beta < 0$ as well, the corresponding $\chi^2_{min}$ is 6.28 ( Note that this small value of $\chi^2_{min}$ is due to the large error bars that we have in  current growth data). In terms of likelihood value, this difference is negligible. Hence both eternal and transient accelerating models are allowed by the growth data with equal likelihood and at present there is no way one can distinguish between the two. With future experiments having tighter constraints on growth history, we hope that a transient acceleration may be distinguished from an eternal acceleration with sufficient confidence level.

In conclusion, we propose a GCG parametrization for the growth history of the Universe and tested it against variety of dark energy models within the framework of Einstein gravity. In future,  we intend to test this parametrization in modified theories as well as in scalar tensor theories of gravity.

\section{Acknowledgement}

SS acknowledges DST, India for the financial support through the FastTrack Scheme (SR/FTP/PS-104/2010). GG acknowledges the Senior Research Fellowship provided by the Council of Scientific and Industrial Research, Govt. of India. SS also acknowledges the hospitality  of IUCAA, India where a 
part of the work was done. AAS acknowledges the financial support from SERC, DST, Govt. of India through the project grant (Grant no: SR/S2/HEP-43 (2009)). 


\end{document}